\documentclass[aps,prl,preprint,superscriptaddress]{revtex4}

\RequirePackage{lineno}
\usepackage{amsmath,bm}
\usepackage{mathrsfs}
\usepackage{amsfonts}
\usepackage{graphicx}
\usepackage[normalem]{ulem}
\usepackage{epstopdf}
\usepackage{dcolumn}
\usepackage{amsmath}
\usepackage{epsfig}
\usepackage{indentfirst}
\usepackage{psfrag}
\usepackage{subfigure}
\usepackage{amssymb}
\usepackage{color}
\usepackage{setspace}
\usepackage{multirow}

\def\BiPd{$\beta$-Bi$_2$Pd}

\newcommand{\nacho}[1]{{\color{black} #1}}

\makeatletter
\newcommand\colorsout[1]{\bgroup \markoverwith{\textcolor{#1}{\rule[0.5ex]{2pt}{0.4pt}}}\ULon}

\makeatother

\begin{document}
	\title{Probing  magnetic interactions between Cr adatoms on the $\beta$-Bi$_2$Pd superconductor}

	\author{Deung-Jang Choi}
	\affiliation{CIC nanoGUNE,  20018 Donostia-San Sebasti\'an, Spain}
	\affiliation{Centro de F{\'{\i}}sica de Materiales
	CFM/MPC (CSIC-UPV/EHU),  20018 Donostia-San Sebasti\'an, Spain}
	\affiliation{Donostia International Physics Center (DIPC),  20018 Donostia-San Sebasti\'an, Spain}
	
	\author{Carlos Garc\'{\i}a Fern\'andez}
	\affiliation{Donostia International Physics Center (DIPC),  20018 Donostia-San Sebasti\'an, Spain}
	
	\author{Edwin Herrera}
	\affiliation{Laboratorio de Bajas Temperaturas, Departamento de F\'{\i}sica de la Materia Condensada, Instituto Nicol\'as Cabrera and Condensed Matter Physics Center (IFIMAC), Universidad Aut\'onoma de Madrid, E-28049 Madrid, Spain}
	
	\author{Carmen Rubio-Verd{\'u}}
	\affiliation{CIC nanoGUNE,  20018 Donostia-San Sebasti\'an, Spain}
	
	\author{Miguel M. Ugeda}
	\affiliation{CIC nanoGUNE, 20018 Donostia-San Sebasti\'an, Spain}
	\affiliation{Ikerbasque, Basque Foundation for Science, 48013 Bilbao, Spain}
	
	\author{Isabel Guillam{\'o}n}
	\affiliation{Laboratorio de Bajas Temperaturas, Departamento de F\'{\i}sica de la Materia Condensada, Instituto Nicol\'as Cabrera and Condensed Matter Physics Center (IFIMAC), Universidad Aut\'onoma de Madrid, E-28049 Madrid, Spain}
	
	\author{Hermann Suderow}
	\affiliation{Laboratorio de Bajas Temperaturas, Departamento de F\'{\i}sica de la Materia Condensada, Instituto Nicol\'as Cabrera and Condensed Matter Physics Center (IFIMAC), Universidad Aut\'onoma de Madrid, E-28049 Madrid, Spain}
	
	\author{Jos\'{e} Ignacio Pascual}
	\affiliation{CIC nanoGUNE, 20018 Donostia-San Sebasti\'an, Spain}
	\affiliation{Ikerbasque, Basque Foundation for Science, 48013 Bilbao, Spain}
	
	\author{Nicol{\'a}s Lorente}
	\affiliation{Centro de F{\'{\i}}sica de Materiales
		CFM/MPC (CSIC-UPV/EHU),  20018 Donostia-San Sebasti\'an, Spain}
	\affiliation{Donostia International Physics Center (DIPC),  20018 Donostia-San Sebasti\'an, Spain}
	
	\begin{abstract}
We show that the magnetic ordering of coupled atomic dimers on a superconductor is revealed by their intra-gap spectral features. Chromium atoms on the superconductor $\beta$-Bi$_2$Pd surface display Yu-Shiba-Rusinov bound states, detected as pairs of intra-gap excitations in the tunneling spectra. We formed Cr dimers by atomic manipulation and found that their intra-gap features appear either shifted or split with respect to single atoms. The spectral variations reveal that the magnetic coupling of the dimer changes between ferromagnetic and antiferromagnetic depending on its disposition on the surface, in good agreement with density functional theory simulations. These results prove that superconducting intra-gap state spectroscopy is an accurate tool to detect the magnetic ordering of atomic scale structures.

	\end{abstract} \date{\today}

	\maketitle % \linenumbers

\nacho{Magnetic impurities have} a detrimental effect on superconductivity \cite{Matthias1958}. \nacho{The exchange interaction between the atomic magnetic moments  and  Cooper pairs produces localized bound states inside the superconducting (SC) gap, known as Yu-Shiba-Rusinov~\cite{Yu,Shiba,Rusinov} states (Shiba states in the following)}. \nacho{Previous} scanning tunneling spectroscopy measurements found
that Shiba states can be detected in \nacho{magnetic adatoms} on a superconducting surface
\cite{Yazdani_1997} as very sharp intra-gap peaks in \nacho{their quasiparticle excitation} spectra \cite{Ji_2008,Ruby_2015}.  \nacho{The position of the peaks inside the gap is very sensitive to the exchange interaction of the impurity with the surface \cite{Franke_2011}, whereas their number can be related to the amount of spin-polarized atomic orbitals  \cite{Ruby2016,Choi_2017} and the spin's multiplet in the presence of anisotropy \cite{Hatter_2015}.}

\nacho{In spite of the local character of the atomic scattering potential, the amplitude of Shiba wavefunctions can  extend over several nanometers \cite{Gerbold_2015,Ruby2016} and interfere with other Shiba states of close-by impurities \cite{Ji_2008,Liljeroth2017}. 
It has been predicted that the sub-gap spectral features of a pair of interacting impurities depend on the alignment of their magnetic moments \cite{Flatte_2000}.  
Hybridization of Shiba states in atomic-scale structures is a prerequisite for the formation of extended Shiba-bands  \cite{Ruby_2015b}, which
under certain circumstances may develop exotic properties such as
topological superconductivity  \cite{vonOppen,Nadj-Perge_2014}. Therefore,
examining the character of Shiba-hybridization at the atomic scale and its relation with the magnetic alignment of the impurities is crucial. }

\nacho{Here we study quasiparticle excitation spectra of Cr atomic dimers on the surface of the superconductor \BiPd\ and show that they reveal the alignment of their magnetic moments. }  We use the tip of a scanning
tunneling microscope (STM) to manipulate individual Cr atoms and form
atomically precise dimers with different interatomic orientations
and distances.  
Differential conductance measurements on the Cr dimers
reveal that their Shiba states shift or
split with respect to those on isolated atoms depending on the dimer's arrangement on the surface. Furthermore, the spatial distribution of the hybridized Shiba states 
resemble either anti-bonding or bonding components of a "Shiba molecule"  \cite{Flatte_2000,Morr2006,Yao2014,Kim2015}.
With the support
of density functional theory (DFT) simulations, we conclude that the
evolution of Shiba states in each case is originated by either the
antiparallel or parallel alignment of the dimer's atomic magnetic moments.

 %--
 % STM manipulations of Cr on BiPd
 %--
 %-----------------%
 % Figure 1
 %-----------------%
 \begin{figure*}[!ht]
 	\begin{center}
 		\includegraphics[width=.85\linewidth]{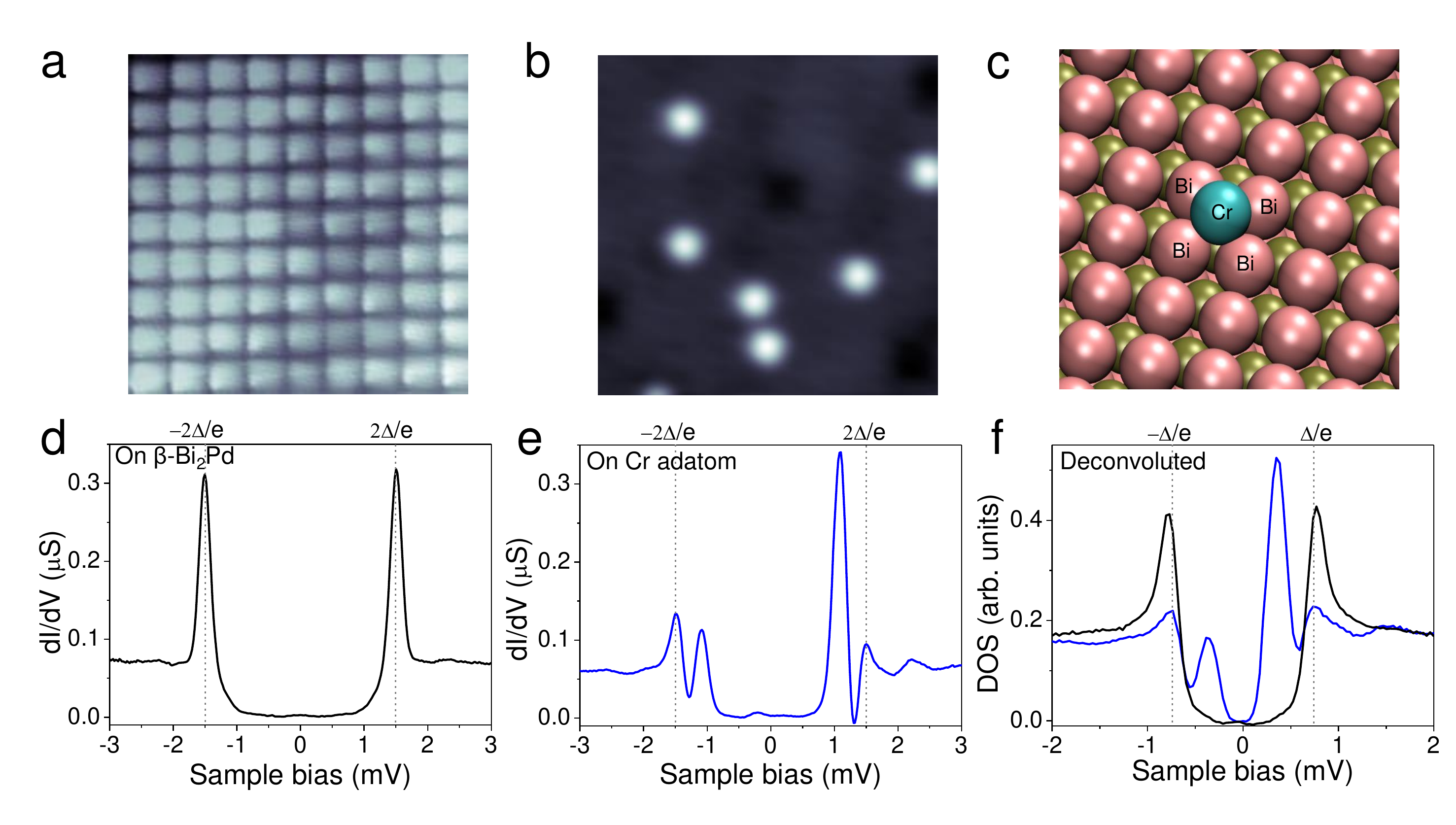}
 	\end{center}
 	\vspace{-0cm}
 	\caption{
		Cr atoms on the Bi-terminated $\beta$-Bi$_2$Pd surface.
		(a) Atomic resolution of a Bi-terminated clean
		$\beta$-Bi$_2$Pd surface in a constant height mode with the junction resistance of 15$k\Omega$ (size: $3 \times3$ nm$^2$). (b)
		STM image of Cr atoms deposited on the Bi-terminated
		$\beta$-Bi$_2$Pd surface (V = 1~V, I = 10~pA, size: $10
		\times10$ nm$^2$).  (c) Results of a full DFT relaxation
		showing that the most stable position of a Cr atom
		(green) is at hollow site (Bi atoms in pink and Pd atoms
		in bronze).  Differential-conductance spectra measured
		(d) on the bare surface and (e) on a Cr adatom,  using a superconducting
		$\beta$-Bi$_2$Pd tip. Tunneling between
		coherent peaks of tip and sample results in sharp dI/dV
		peaks at $\pm 2\Delta/e$ (dashed lines in the plots).
		(f) Density of states obtained by deconvoluting the
		tip DOS from the the spectra  on the Cr adatom (blue)
		and on the bare surface (black). Analysis of STM and STS data was performed with the WSxM \cite{Wsxm}
		and SpectraFox \cite{Spectrafox} software packages.}
 	\label{Fig1} 
 \end{figure*} 
 %-----------------%

Our experiments were carried out on the Bi-terminated surface of a \BiPd\
crystal (see Methods in Supplementary Information (SI) \cite{SI}), cleaved under  ultra-high-vacuum
(UHV) conditions, and subsequently transferred \textit{in situ} into a
cryogenic  STM for  measurements at T=1.2~K.  The exposed  surface is
atomically clean and shows a squared atomic  structure (Fig.~\ref{Fig1}a), with
periodicity in agreement with the \BiPd\ unit cell (lattice parameter
$a$ = 3.36~\AA).  Next, we evaporated small amounts of Cr atoms on the
pristine surface at low temperature (T = 15~K) to obtain Cr densities
similar to that shown in the STM image of Fig.~\ref{Fig1}b. Individual
Cr adatoms appear as protrusions 110~pm high,  absorbed on hollow sites of
the bismuth surface. This agrees with the minimum energy  configuration obtained from Density Functional Theory (DFT) simulations, shown in  Fig.~\ref{Fig1}c and in the SI \cite{SI}.

\nacho{The  stoichiometric compound  \BiPd\ is a s-wave superconductor with a single gap of magnitude $\Delta=0.76~$meV~\cite{Herrera_2015}. The superconducting properties are very isotropic \cite{Kacmarcik2016}, leading to a narrow gap distribution of just a few tens of $\mu$eV wide~\cite{Herrera_2015}, in spite of its square Fermi surface \cite{Iwaya2017,sakano_2015}. } We determined the  LDOS of pristine and Cr-decorated  regions
by means of differential conductance (dI/dV) spectra. 
To enhance the spectral resolution, we employed superconducting STM tips obtained by indenting a W tip into the $\beta$-Bi$_2$Pd surface, which has an isotropic single superconducting gap. 
As a consequence,  the tunneling spectra  result from  the convolution of
tip and sample LDOS~\cite{Ji_2008,Franke_2011,Franke2017}.  The dI/dV
spectrum of the pristine $\beta$-Bi$_2$Pd surface, Fig.~\ref{Fig1}d,
shows a conductance gap with two sharp peaks at $\pm1.52$ mV,
i.e. at $\pm2\Delta /e$, due to tunneling between quasiparticle (QP)  peaks at $\pm\Delta$ in tip and sample.  Typical dI/dV
spectra acquired on top of single Cr adatoms (Fig.~\ref{Fig1}e) show
smaller QP-peaks and a pair of additional peaks inside the spectral gap
(at V=$\pm1.1\pm0.1$~mV) indicating the formation of Shiba states at the
locations of the adsorbed Cr atoms. 
Figure~\ref{Fig1}f compares the LDOS of pristine surface and Cr adatoms
extracted from the tunneling spectra following the deconvolution procedure
of Refs.~\cite{Choi_2017,Pillet_2010}.

% dI/dV data: Shiba states, Shiba dimers
%--

%-----------------%
% Figure 2
%-----------------%
\begin{figure*}[tb]
        \begin{center}
             \includegraphics[width=0.85\linewidth]{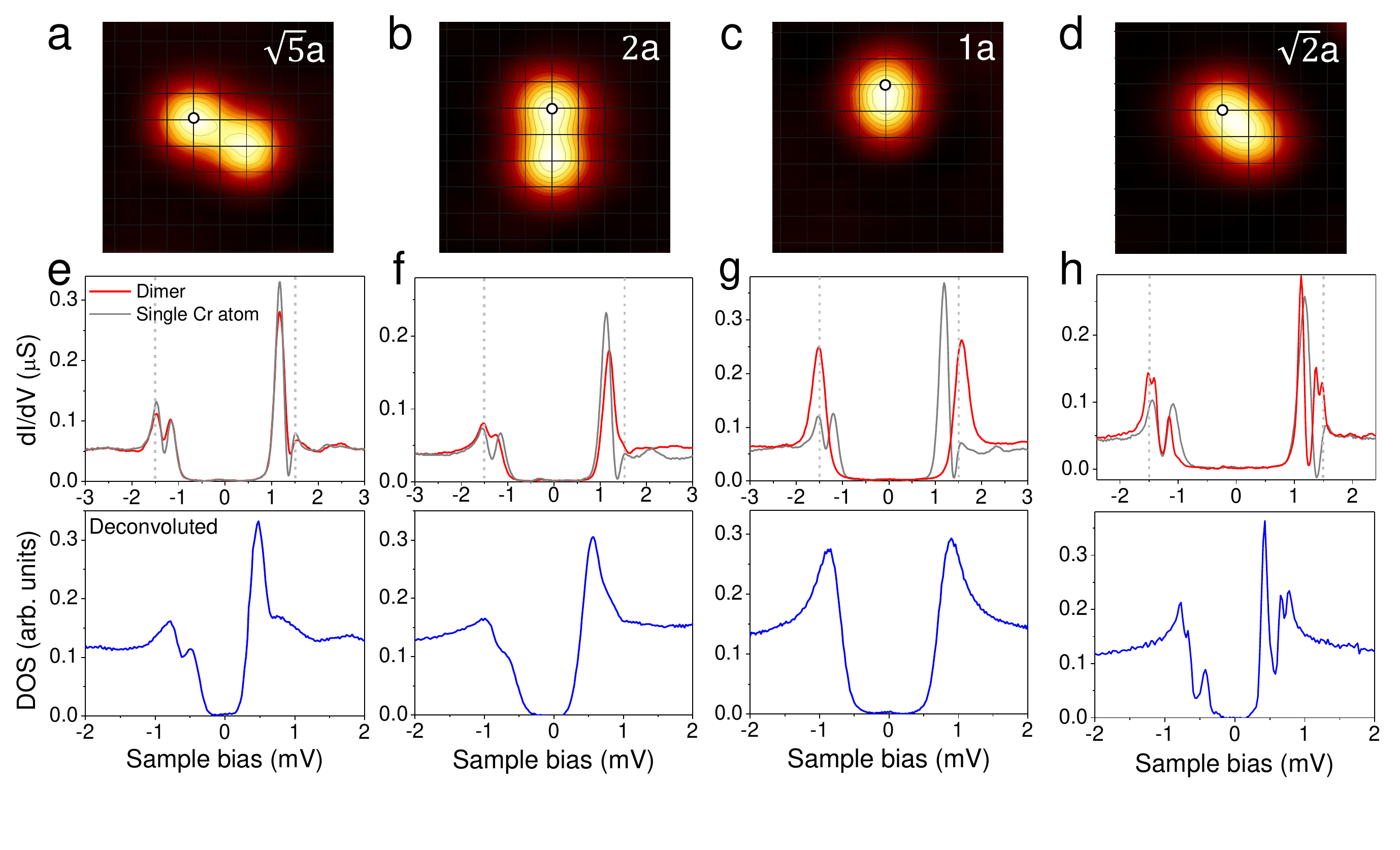}
        \end{center}
        \vspace{0cm}
        \caption {(a)--(d) STM  image of Cr dimers at indicated spacings  (V=1~V and I=10~pA, size:$3 \times 3$ nm$^2$).  The square mesh depicts the surface \BiPd\ lattice. (e)--(h)  Differential conductance spectra on the atom indicated by a white dot  before (gray) and after (red) the second atom is moved to the dimer  position. As in Fig.~\ref{Fig1}, all spectra are measured using  superconducting  tips with the same $\Delta$ as the substrate (dashed lines mark the  $\pm2\Delta/e$ bias). The corresponding tip-deconvoluted DOS is shown in the lower panels. }
        \label{Fig2}
 \end{figure*} %-----------------%

To explore the effect of magnetic interactions on Cr Shiba states,   we studied the evolution of  intra-gap spectral features of Cr atomic dimers constructed by STM lateral manipulation. 
Figs.~\ref{Fig2}a--d show STM images of the four types of dimers that show sizable spectral variations. The atoms are separated by (a)  $d=\sqrt{5}a$, (b) $d=2a$, (c) $d=1a$ and (d) $d=\sqrt{2}a$, corresponding to 
distances between hollow sites. For larger Cr-Cr separations, the dI/dV spectra
are unaffected by the neighboring adatom. In every case, we measured the dI/dV spectrum of
a target Cr adatom first isolated (grey in Figs.~\ref{Fig2}e--h),
and then after a second adatom is precisely positioned nearby (red in
Figs.~\ref{Fig2}e--h). The spectrum on the $d=\sqrt{5}a$ dimer shows only a faint effect of the interaction between atoms   (Fig.~\ref{Fig2}e).  
When their separation is reduced to $d=2a$, the Shiba peaks appear closer to the SC gap edge ($\epsilon \sim$450~$\mu$eV, Fig.~\ref{Fig2}f), overlapping with the QP peaks of the superconductor  and evolving  into broader spectral features. Finally, at very short distances ($d=1a$), the Shiba excitations are absent from the spectrum, which now show symmetric but broader QP peaks  (Fig.~\ref{Fig2}g). The observed tendency is that Shiba peaks shift towards the QP continuum as the interatomic distance decreases.

Interestingly, when the Cr adatoms are moved to contiguous hollow sites along the surface diagonal, $d$=$\sqrt{2}a$, the spectra
show a more complex structure with additional features close to the
QP peaks (Fig.~\ref{Fig2}h). In this case, the DOS is composed of two
intra-gap peaks for particle states and two peaks for hole states with
separation of $\sim$250 $\pm$ 50$\mu$eV. This pattern resembles the
splitting of hybridized Shiba states predicted for atoms interacting
ferromagnetically~\cite{Flatte_2000,Morr2006,Yao2014,Kim2015}.

%Spatial distribution

%-----------------%
% Figure 3
%-----------------%
\begin{figure}[t]
	\includegraphics[width=.65\linewidth]{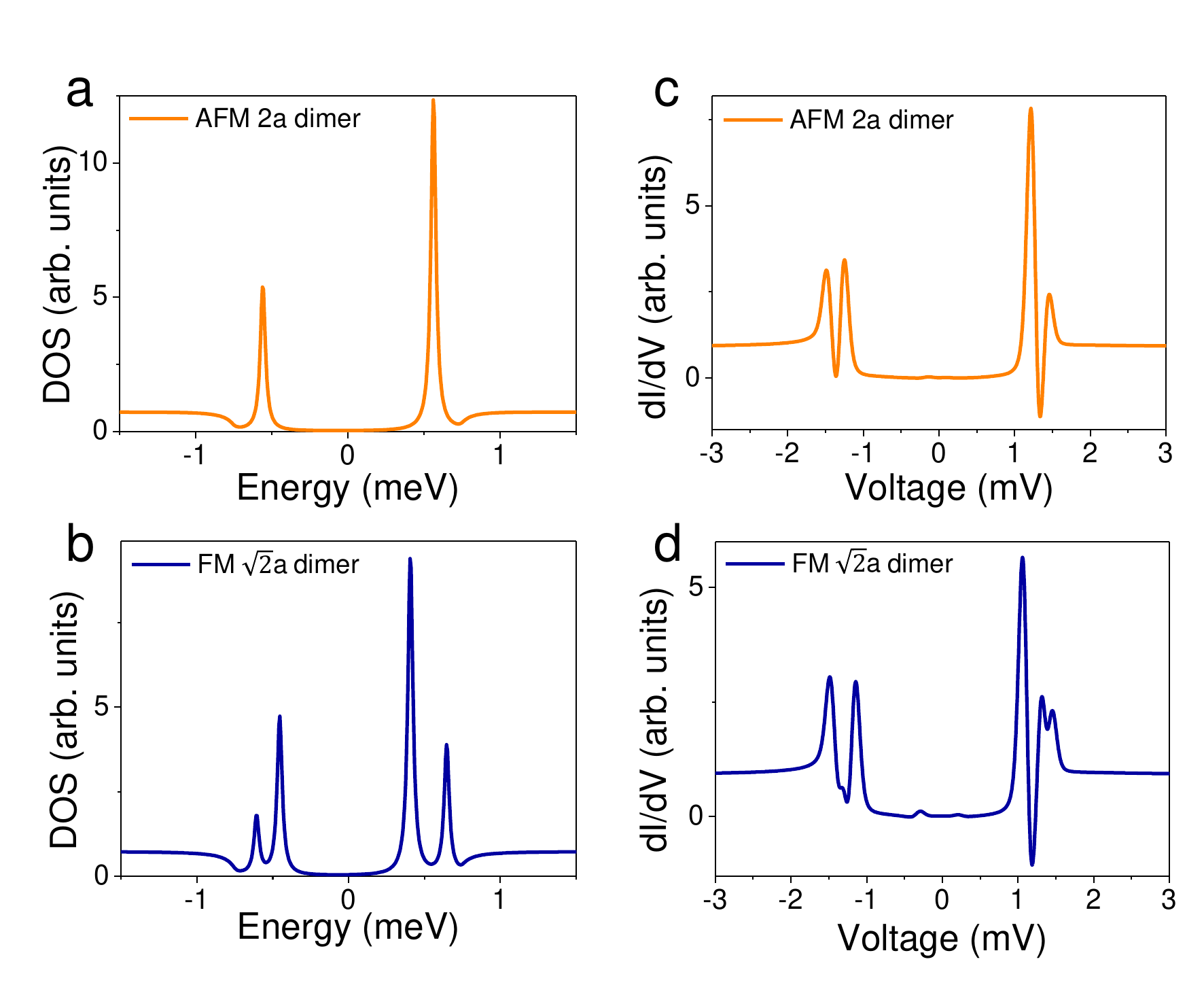}
	\vspace{-0.3cm}
	\caption{ Model  calculation of the Shiba density of states (DOS) of  (a) an  antiferromagnetically ordered dimer separated by d=$2a$ and (b) a ferromagnetically ordered one  at d=$\sqrt{2}a$ (see \cite{SI}). For both dimer structure,  ferromagnetic  order leads to split Shiba peaks,  while AFM shows only a degenerate one.  Panels (c) and (d) shows the corresponding dI/dV spectra obtained  by  convoluting the dimer's DOS with a superconducting tip   DOS.}
 \label{Fig3}
\end{figure}
%-----------------% 

To interpret the observed evolution of intra-gap spectra in terms of the \nacho{magnetic} alignment of the Cr spins, we show in Fig.~\ref{Fig3} the
simulated DOS of a  ferromagnetic (FM) $\sqrt{2}a$ dimer, and the
corresponding dI/dV spectrum, obtained using an extension of Flatt\'e
and Reynolds model~\cite{Flatte_2000} (details in SI~\cite{SI}). The model parametrizes first the  electron-Cr interaction of a Cr monomer on the \BiPd\ substrate to fit their experimental spectral features. Using this parametrization, the model reproduces  a splitting of Shiba states for the  $\sqrt{2}a$  dimer  if the two spins of the Cr atoms are assumed to be parallel. Imposing an antiferromagnetic (AFM) alignment to the magnetic moments of a $2a$ dimer   leads instead to a single Shiba peak (Fig.~\ref{Fig3}a), as observed in the experiments. The model further captures the additional role of potential scattering in the subgap states, which
induces a small asymmetry in particle and hole energy of
the states for both AFM and FM cases.    
These results thus confirm that interacting Shiba states  of Cr atoms can evolve into  molecular-like states with  bonding and antibonding configurations  depending on the type of magnetic order.

%Spatial distribution

%-----------------%
% Figure 4
%-----------------%
\begin{figure}[t]
 		\includegraphics[width=.85\linewidth]{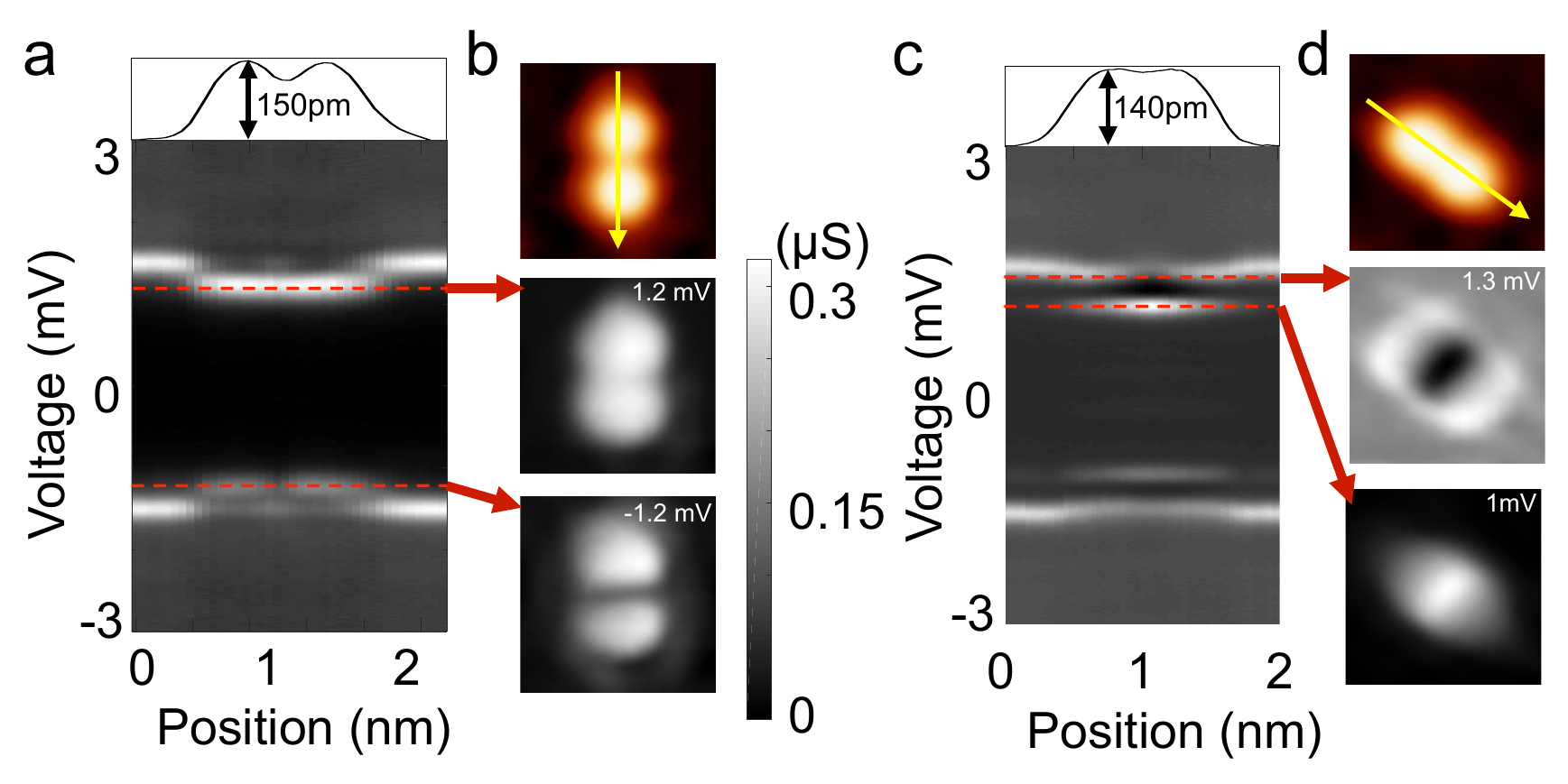}
 	\vspace{-0.3cm}
	\caption{ Topography profiles and spectral (dI/dV vs. V) maps measured along the axes of the (a) $2a$
	and (c)  $\sqrt{2}a$ dimers (gray-scale covers dI/dV range from 0 to 0.3 $\mu$S). Top panels in (b) and (d) correspond to the dimer STM image, scanned at 3~meV and  10~pA. Below, we show the 2D spatial distribution of
	the differential conductance at the bias of the Shiba peaks for (b) the $2a$ dimer (size: $3 \times 3$ nm$^2$)
	and (d) the $\sqrt{2}a$ dimer (size: $2.5 \times 2.5$ nm$^2$), at the
	given energies.} 
\label{Fig4}
\end{figure} %-----------------%

The different type of Shiba hybridization is reflected in their respective  spatial distribution of bound-state excitation amplitude \cite{Flatte_2000}.  Figure \ref{Fig4}a and \ref{Fig4}c show
the spectral maps (i.e. dI/dV vs distance and bias) measured along the
axis of the $2a$ and $\sqrt{2}a$  dimers, respectively. In all cases,
the amplitude of the QP  peaks decreases substantially over the
dimer and a distinct pattern of sub-gap excitations emerges for every
atomic arrangement. For the $2a$ dimer, the amplitude of the Shiba peaks
is clearly maximum over each atom and shows a nodal plane in between,
both for  particle and hole components (Fig. \ref{Fig4}b). This is in
good agreement with predictions for the AFM case~\cite{Flatte_2000},
picturing the  hybridized Shiba states as degenerate states localized
around each of the two impurities.  In contrast, the $\sqrt{2}a$
dimer shows a  different pattern for each  of the split Shiba states.
The state with lower energy appears with more amplitude in the region
between the two atoms, while the higher one shows a nodal plane between
them (see Fig.\ref{Fig4}c). This is clearly visualized in dI/dV spatial
maps at each excitation energy (Fig. \ref{Fig4}d), showing a  bonding and
anti-bonding  pattern for the lower and higher Shiba states, respectively.

\begin{figure}[th]
  	\begin{center}
  		\includegraphics[width=.95\linewidth]{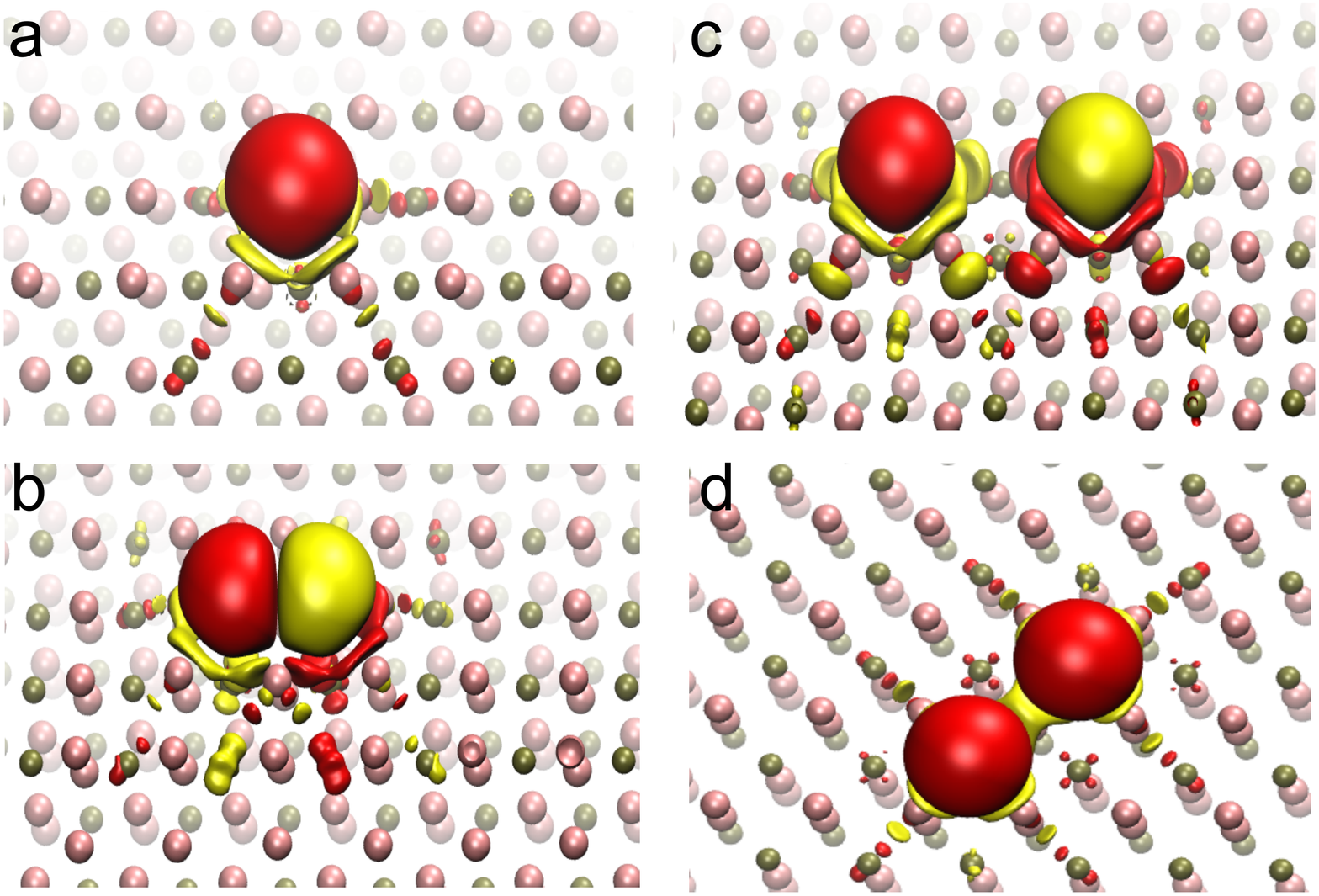}
  	\end{center}
  	\vspace{-0.3cm}
  	\caption{ Spin polarization (a) of a Cr adatom and of various dimers (d=1$a$ (b),
 2$a$ (c) $\sqrt{2}a$ (d))  on the \BiPd\ surface. In each case, the
 results correspond to the minimum-energy configuration obtained from
 DFT simulations of the relaxed system. The plot shows 3D isosurfaces
 of constant electronic density difference between majority and minority
 spins, in the normal state. The surface is tilted differently in every
 case to picture the extension of the spin density to inner layers. Bi
 atoms are represented in pink and    Pd atoms in bronze. The isocontours
 of spin are yellow and red for each of the two spin components (the
 isocontour is 0.005 e/\AA$^3$ in all graphs).} 
\label{Fig5}
  \end{figure}

Density function theory further confirms the magnetic ordering of the adsorbed dimers. We first calculated the
adsorption properties of a single Cr atom on \BiPd\ .  
The calculations find that
Cr adatoms preferentially
adsorb on the hollow site, with a subsurface Pd atom directly underneath, as shown
in Fig.~\ref{Fig1}c. As a result of the bonding interaction, 
the adatom's next-neighbor Pd and Bi
atoms  slightly approach towards the Cr adatom  by 10 pm
and 5 pm, respectively inducing a small distortion of the surface (see SOM). The Cr adatom lies 1.59~\AA~above
the  surface layer and keeps a total magnetic moment of 4.4 $\mu_B$,
close to the 5 $\mu_B$  of a free atom. The large spin polarization of
the system can be pictured by isosurfaces   of the electronic
density difference between majority and minority spins, shown in
Fig.~\ref{Fig5}a. The Cr atom  polarizes the
four nearest-neighbour Bi atoms antiferromagnetically,  and   the second-layer
Pd atoms ferromagnetically.

%-- 
% Cr DFT results 
%--

  %-----------------%
  % Figure 5
  %-----------------%
   %-----------------%

Next, we calculated the adsorption properties of the different types of dimers
found in the experiment. The most stable dimer configuration on the surface 
has both atoms absorbed on contiguous hollow sites separated by a distance d=$1a$. 
This corresponds to the compact dimer shown in Fig.~\ref{Fig2}c and Fig.~\ref{Fig5}b. The two atoms interact strongly and approach,  reducing the  Cr-Cr distance by 80 pm (from $a=$3.46~\AA\ to d=2.74~\AA).  
The Cr atoms are clearly antiferromagnetically coupled with an energy 152 meV lower than the  ferromagnetic configuration
(E$_{AFM}$-E$_{FM}$=$\Delta$E=-152 meV).  The $2a$ dimer shows a much
weaker interatomic interaction, and each Cr appears with a negligible
deviation from the single adatom adsorption geometry. Their atomic
spins interact indirectly via the substrate with a slight preference
for  antiferromagnetic  ordering ($\Delta$E=-10 meV), in agreement with
the assignment from the experiments.  The spin density of the $2a$
configuration (Fig.~\ref{Fig5}c) pictures this  antiferromagnetic  arrangement in the ground state. It also shows that the intermediate Bi atoms have  a stronger magnetic polarization, suggesting that these atoms might mediate the interaction between Cr adatoms.

The dimer along  neighbouring diagonal sites (d= $\sqrt{2}a$
in Fig.~\ref{Fig2}d and Fig.~\ref{Fig5}d) shows a clear preference for a  ferromagnetic  ground-state, with $\Delta$E=19 meV, again  in good agreement with the
interpretation of Shiba spectral features. Interestingly, both Cr atoms appear
connected via a single Bi atom (Fig.~\ref{Fig5}d), which shows opposite magnetization and, probably, forces the ferromagnetic ordering of the dimer.

The DFT results thus confirm the magnetic  ordering  deduced
from the analysis of Shiba states. Both $1a$ and $2a$ dimers are
antiferromagnetically aligned, in agreement with the measurement of a single pair of intra-gap excitations. The observed shift of the Shiba excitations towards the gap edge  with reducing Cr--Cr distance  reflects
the  weakening of the bound state's energy, probably due to the reduction of the total magnetization of  antiferromagnetically  coupled dimers.  For the $\sqrt{2}a$ dimer, the splitting  of Shiba states and their peculiar bonding-antibonding  spatial distribution   
reflect the  ferromagnetic   coupling of the dimer.  %Inclusion of spin-orbit coupling does not   lead to non-collinear ordering for atomic dimers (see SOM).

%-- % Model interaction %--

%--
% Summary and Conclusions
%--

In summary, we demonstrated that the spectral features of coupled  Shiba
states reflect the magnetic ordering of interacting atoms. We proved this
by studying dimers of  Cr atoms on a superconducting  $\beta$-Bi$_2$Pd
surface, assembled by atomic manipulation using a low-temperature
STM. Differential conductance spectra reveal sub-gap excitations
associated to Shiba states, which evolve as the atoms are brought
to proximity, reflecting their  mutual
spin alignment. We found that
different atomic arrangements on the surface result in   shifts or splits   of the atomic Shiba features. Furthermore, the  spatial distribution
of the  Shiba peaks  for the $\sqrt{2}a$ dimer resembles bonding and antibonding states, as
predicted for a ferromagnetic dimer.  
DFT simulations confirm  the magnetic ordering deduced 
from the spectra analysis. Therefore, the measurement of  Shiba excitation spectra  of magnetically
coupled  impurities  is an excellent probe of  their  magnetic ordering.

We thank Javier Zaldivar and Joeri de Bruijckere for developing the deconvolution process and Sebastian Bergeret for discussions. DJC and JIP thank the
European Union for support under the H2020-MSCA-IF-2014 Marie-Curie
Individual Fellowship program (proposal number 654469), Spanish MINECO (MAT2016-78293-C6-1-R), Diputacion Foral de Gipuzkoa for grant N$^\circ$
64/15, and the European Regional Development Fund (ERDF).
NL thanks Spanish MINECO (Grants No. MAT2015-66888-C3-2-R). MMU acknowledges spanish MINECO (MAT2014-60996-R).
EW, IG and HS acknowledge FIS2014-54498-R and MDM-2014-0377, 
ERC PNICTEYES Grant Agreements No. 679080, Departamento Administrativo de Ciencia, Tecnolog\'{\i}a e Innovaci\'on, COLCIENCIAS (Colombia), Programa doctorados en el exterior, convocatoria 568-2012 
Comunidad de Madrid through program Nanofrontmag-CM
(Grant No. S2013/MIT-2850).

\bibliography{dimer}

\end{document}

\title{Probing  magnetic interactions between Cr adatoms on the $\beta$-Bi$_2$Pd superconductor}

\author{Deung-Jang Choi}
\affiliation{CIC nanoGUNE,  20018 Donostia-San Sebastian, Spain}
\affiliation{Centro de F{\'{\i}}sica de Materiales
	CFM/MPC (CSIC-UPV/EHU),  20018 Donostia-San Sebasti\'an, Spain}
\affiliation{Donostia International Physics Center (DIPC),  20018 Donostia-San Sebasti\'an, Spain}

\author{Carlos Garc\'{\i}a Fern\'andez}
\affiliation{Donostia International Physics Center (DIPC),  20018 Donostia-San Sebasti\'an, Spain}

\author{Edwin Herrera}
\affiliation{Laboratorio de Bajas Temperaturas, Departamento de F\'{\i}sica de la Materia Condensada, Instituto Nicol\'as Cabrera and Condensed Matter Physics Center (IFIMAC), Universidad Aut\'onoma de Madrid, E-28049 Madrid, Spain}

\author{Carmen Rubio-Verd{\'u}}
\affiliation{CIC nanoGUNE,  20018 Donostia-San Sebastian, Spain}

\author{Miguel Moreno Ugeda}
\affiliation{CIC nanoGUNE, 20018 Donostia-San Sebastian, Spain}
\affiliation{Ikerbasque, Basque Foundation for Science, 48013 Bilbao, Spain}

\author{Isabel Guillam{\'o}n}
\affiliation{Laboratorio de Bajas Temperaturas, Departamento de F\'{\i}sica de la Materia Condensada, Instituto Nicol\'as Cabrera and Condensed Matter Physics Center (IFIMAC), Universidad Aut\'onoma de Madrid, E-28049 Madrid, Spain}

\author{Hermann Suderow}
\affiliation{Laboratorio de Bajas Temperaturas, Departamento de F\'{\i}sica de la Materia Condensada, Instituto Nicol\'as Cabrera and Condensed Matter Physics Center (IFIMAC), Universidad Aut\'onoma de Madrid, E-28049 Madrid, Spain}

\author{Jos\'{e} Ignacio Pascual}
\affiliation{CIC nanoGUNE, 20018 Donostia-San Sebastian, Spain}
\affiliation{Ikerbasque, Basque Foundation for Science, 48013 Bilbao, Spain}

\author{Nicol{\'a}s Lorente}
\affiliation{Centro de F{\'{\i}}sica de Materiales
	CFM/MPC (CSIC-UPV/EHU),  20018 Donostia-San Sebasti\'an, Spain}
\affiliation{Donostia International Physics Center (DIPC),  20018 Donostia-San Sebasti\'an, Spain}

\renewcommand{\abstractname}{\vspace{3cm} }	
\date{\today}

\begin{abstract}
 	\setstretch{1}
	\tableofcontents		
\end{abstract}	

\maketitle
\setstretch{1.0}

\newpage
\subsection{Note 1 - Sample preparation and the spectra of vacancies and defects on \mbox{$\beta$-Bi$_2$Pd} surfaces}

The experiments were performed on $\beta$-Bi$_2$Pd crystals  grown from high-purity Bi (Alfa Aesar 99.99$\%$) and Pd (Alfa Aesar 99.95$\%$) following the procedure described
in Ref.~\cite{Herrera_2015}.   STM/STS
measurements were carried out in an UHV low-temperature Joule-Thomson
STM at a base temperature of 1.15 K.  The differential conductance was
measured using lock-in detection with a modulation of 40$\mu$V rms at 938.6 Hz. Analysis of STM and STS data was performed with the WsXm \cite{Wsxm}
and SpectraFox \cite{Spectrafox} (http://www.spectrafox.com) software packages. 

To increase the energy resolution beyond the limit imposed by the Fermi-Dirac distribution of a metallic tip, we use in our experiments superconducting tips prepared by gently dipping a W-tip into the $\beta$-Bi$_2$Pd surface.  This process was repeated until sharp dI/dV peaks due to tunneling between coherent peaks of tip and sample appeared separated by a  4$\Delta/e$  bias gap ($\Delta$=0.75 meV is the order parameter of $\beta$-Bi$_2$Pd)  in the tunneling  spectra of the bare $\beta$-Bi$_2$Pd surface. 
Although a superconducting tip with the bulk order parameter was easily obtained, the energy resolution achieved by these tips varies depending on the experiment.

\begin{figure}[!ht]
	\begin{minipage}[c]{0.70\textwidth}
		\includegraphics[width=0.99\textwidth]{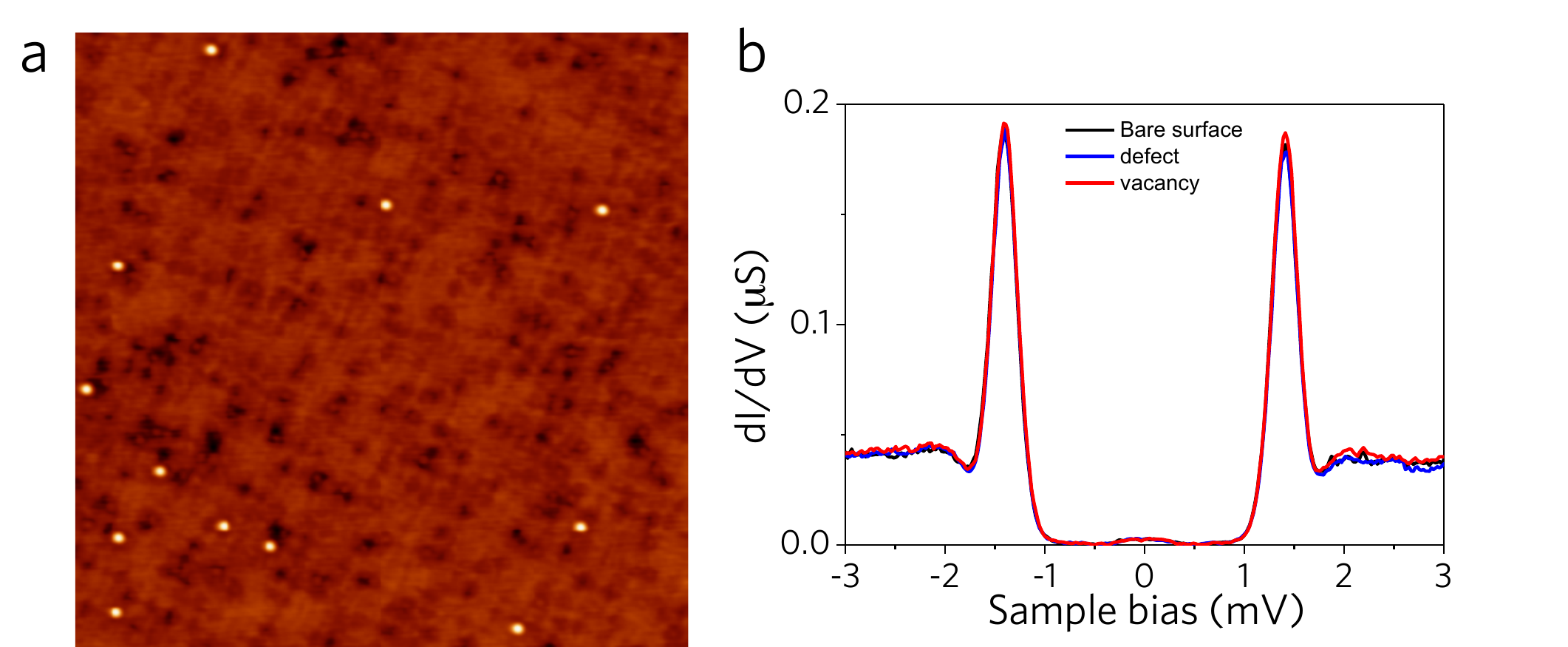}
	\end{minipage}\hfill
        \begin{minipage}[c]{0.30\textwidth}
         \caption{      	Bare surface of the $\beta$-Bi$_2$Pd surface after the \textit{in-situ}
        	exfoliation. (a) Topographic image over $50 \times 50$ nm$^2$ (scanned
        	at 100mV, 10pA).  (b) The differential conductance measured on the defect
        	and the vacancy compared with the one on the bare surface.
        	\label{FigS3}}
        \end{minipage}\hfill
       
\end{figure}

In Fig.~\ref{FigS3}(a) we show an STM image of the pristine $\beta$-Bi$_2$Pd surface.  Atomic-scale protrusions and vacancies (seeing as depressions) were frequently observed over  a $50 \times 50$ nm$^2$ scanned area.  As shown in Fig.~\ref{FigS3}(b), on both types we measure identical tunneling spectra as on the bare surface regions. From this, we conclude that there is no magnetism associated with the adatoms or the vacancies. This contrast with the spectra on deposited Cr atoms, which shows a very characteristic fingerprint, as we show in the main text. Since our samples are cleaved  in ultra-high vacucum conditions, the protrusion cannot be related to contaminants. Instead, it is probable the protrusions are Bi adatoms extracted from the observed vacancies on the Bi surface layer during cleaving.

%-----------------%
% Figure (bare_Bi2Pd)
%-----------------%
%-----------------% 

\newpage
\subsection{Note 2 - Hamiltonian model for Yu-Shiba-Rusinov states of a chain of atoms}

%taken from the method section
The Shiba-chain state and conductance simulation were performed solving
the BCS Hamiltonian in real space in the presence of $\delta$-like exchange, $J_i$,
and scattering, $K_i$, potentials following Refs.~\cite{Flatte_2000,Flatte_1997}.
The model Hamiltonian can then be written as:
\begin{equation}
\hat{H}=\hat{H}^0_{BCS} + \sum^{N_{at}}_{i=1} (J_i \vec{S}_i \cdot \vec{\alpha}
+K_i \tau_3)
\delta (\vec{r}-\vec{r}_i)  .
\label{hamiltonian}
\end{equation}
Here $N_{at}$ is the number of atoms of the Shiba chain and $i$ is the
index for each atom of the chain.
The Pauli matrices for the spin ($\vec{\sigma}=(\sigma_1,\sigma_2, \sigma_3)$) 
and electron-hole ($\vec{\tau}=(\tau_1,\tau_2, \tau_3)$)
sectors are explicitly considered in $\vec{\alpha}$ as~\cite{Shiba}:
\[\vec{\alpha}
=
\frac{\hat{1}_\tau+\hat{\tau_3}}{2}\otimes \vec{\sigma} + \frac{\hat{1}_\tau-\hat{\tau_3}}{2}\otimes \sigma_2 \vec{\sigma} \sigma_2 \]
corresponding to the Nambu spinor~\cite{Shiba}:
\begin{equation}
\Psi (\vec{r})=
\begin{pmatrix}
\psi_\uparrow (\vec{r})\\
\psi_\downarrow (\vec{r})\\
\psi_\uparrow^\dagger (\vec{r})\\
\psi_\downarrow^\dagger (\vec{r})
\end{pmatrix}
\label{Nambu}
\end{equation}
To represent the $\beta$-Bi$_2$Pd superconductor we used for simplicity a single-band BCS Hamiltonian $\hat{H}^0_{BCS}$ with the largest Fermi velocity ($v_F=0.4\times10^{6}$ m/s), and  with  gap $\Delta=0.76$ meV,
as determined in Ref.~\cite{Herrera_2015} and \cite{Kacmarcik2016}.

The resolvent or Green's function of the BCS Hamiltonian 
can be written as a function of the distance $r$ of two points, $r=|\vec{r}-\vec{r}'|$, in
the translationally-invariant superconductor as~\cite{Flatte_1997}:
\small 
\begin{center}
\begin{eqnarray}
G^0_{BCS} (\vec{r},\vec{r}', \omega)
=- \frac{\pi N_0}{k_F r} e^{-\frac{\sqrt{\Delta^2-\omega^2}}{\pi \xi \Delta} r}
\times
\;\;\;\;\;\;\;\;\;\;\;\;\;\;\;\;\;\;\;\;\;\;\;\;\;\;\;\;\;\;\;\;\;\;\;\;\;\;\;\;\;\;\;\;\;\;\;\; &\;&\nonumber \\
& & \nonumber \\
\begin{pmatrix}
\frac{\omega}{\sqrt{\Delta^2-\omega^2}}sin k_F r + cos k_F r &
0 & 0 & -\frac{\Delta}{\sqrt{\Delta^2-\omega^2}} sin k_F r \\
0 & \frac{\omega}{\sqrt{\Delta^2-\omega^2}}sin k_F r - cos k_F r &
\frac{\Delta}{\sqrt{\Delta^2-\omega^2}}sin k_F r & 0 \\
0 & \frac{\Delta}{\sqrt{\Delta^2-\omega^2}}sin k_F r  &
\frac{\omega}{\sqrt{\Delta^2-\omega^2}}sin k_F r + cos k_F r & 0 \\
-\frac{\Delta}{\sqrt{\Delta^2-\omega^2}}sin k_F r & 0 & 0 &
\frac{\omega}{\sqrt{\Delta^2-\omega^2}}sin k_F r - cos k_F r
\end{pmatrix}
\label{G0}
\end{eqnarray}
\end{center} \normalsize
here, $\xi = \hbar v_F / \pi \Delta$ is the coherence length,  $k_F$ the Fermi wave-vector, and $N_0$  the normal metal density at the Fermi energy.
Following the careful study by Vernier~\cite{Vernier}, we see
that the limit $r\rightarrow 0$ is:
\begin{equation}
G^0_{BCS} (\vec{r},\vec{r}, \omega)
= - \pi N_0 Sign [Re(\omega)Im(\omega)]
\begin{pmatrix}
\frac{\omega}{\sqrt{\Delta^2-\omega^2}} & 0 & 0 & -\Delta \\
0 & \frac{\omega}{\sqrt{\Delta^2-\omega^2}} &\Delta& 0\\
0 &\Delta&\frac{\omega}{\sqrt{\Delta^2-\omega^2}}& 0 \\
-\Delta & 0 & 0 & \frac{\omega}{\sqrt{\Delta^2-\omega^2}}
\end{pmatrix}
\end{equation}

The non-collinear Shiba chain problem can be
treated now with a Dyson equation where the self-energy is local
thanks to the $delta$-interaction and given in Nambu-spinor form
by:
\begin{equation}
\Sigma_{i,i}=
\begin{pmatrix}
J_i S_i cos \theta_i + K_i & J_i S_i sin \theta_i e^{-i \phi_i} & 0 & 0 \\
J_i S_i sin \theta_i e^{i \phi_i} & -J_i S_i cos \theta_i + K_i & 0 & 0 \\
0 & 0 & -J_i S_i cos \theta_i - K_i & -J_i S_i sin \theta_i e^{i \phi_i} \\
0 & 0 & -J_i S_i sin \theta_i e^{-i \phi_i} & J_i S_i cos \theta_i - K_i 
\end{pmatrix}
\label{self}
\end{equation}
where each spin $\vec{S}_i$ is now given by its modulus $S_i$ and
two angles $\theta_i$ and $\phi_i$.
The Dyson equation can now be solved in the tight-binding basis
of each atom, $\psi_i (\vec{r})$ and STM
data can be simulated by the local density of states:
\begin{equation}
\rho (\vec{r},\omega) = -\frac{1}{\pi} Im \{ tr[ \sum_{i,j} \psi_i ( \vec{r})
\hat{G}_{i,j} (\omega) \psi_j^* (\vec{r}) \frac{\hat{\tau}_3 + \hat{1}_\tau}{2}]
\}.
\label{rho}
\end{equation}

\subsection{Note 3 - Simulation of Yu-Shiba-Rusinov states and sub gap spectra of Cr dimers}

We have approximated the conductance over dimers of Cr atoms on a $\beta$-Bi$_2$Pd superconductor using a BCS superconductor and two localized magnetic moments (of 5 Bohr magnetons each, according to our DFT simulations in Note 4) at different distances. \nacho{We employed the model described in Note 2, which ignores any direct exchange interaction between Cr atoms. Thus, only the BCS superconductor reacts to the presence of the magnetic moments by the appearance of in-gap bound states. These Shiba states can interact through the BCS superconductor, giving rise to bonding/antibonding  states if the two Cr spins are aligned parallel or to a single bound state for the antiferromagnetic alignment.}
%-----------------%
% Figure (DOS)
%-----------------%
\begin{figure}[tb]
        \begin{center}
                \includegraphics[width=.90\linewidth]{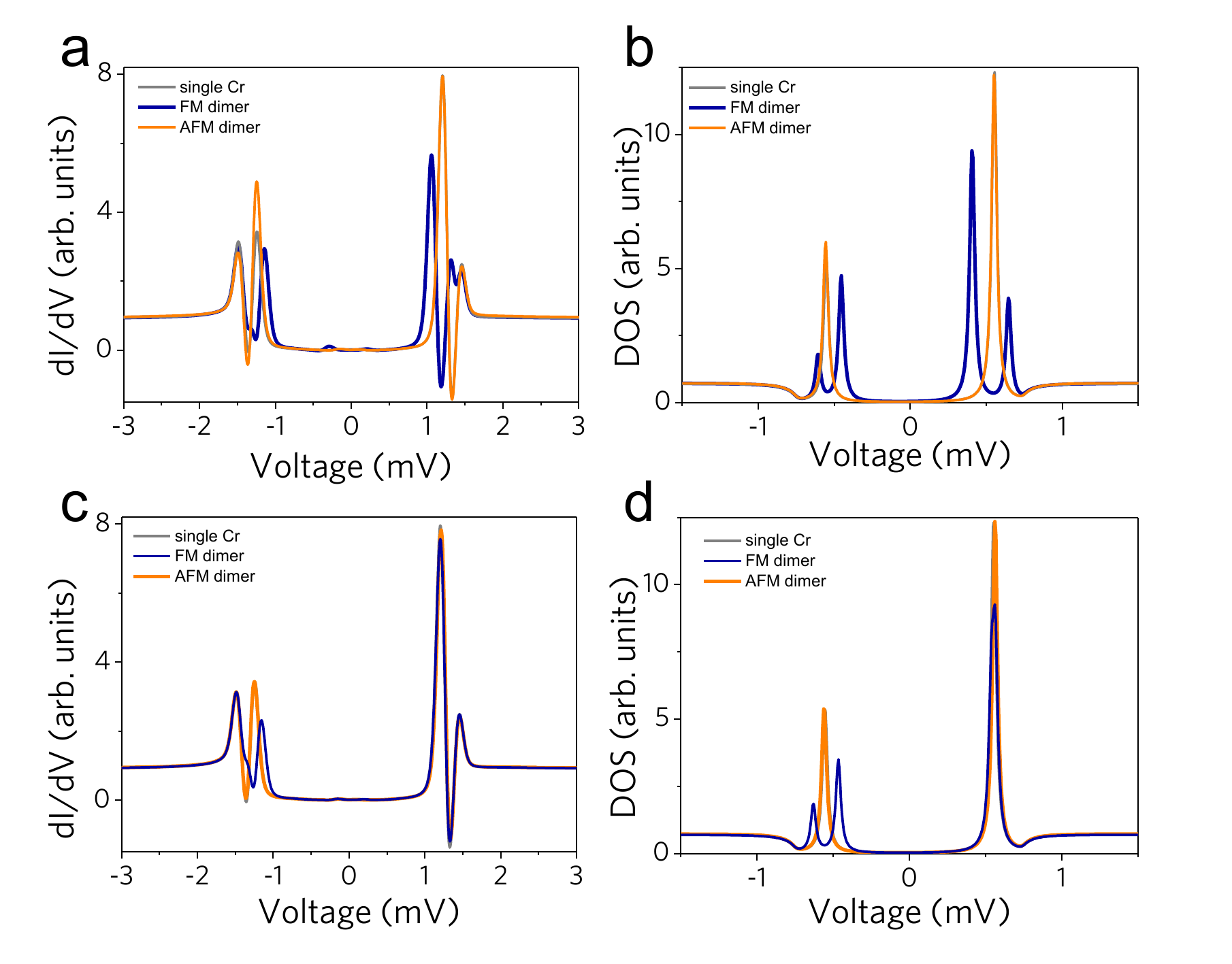}
        \end{center}
        \vspace{-0.3cm}
        \caption{
Simulations of dimers with parallel (FM) or antiparallel (AFM) magnetic moments. Spectra on a Cr atom of (a) $\sqrt{2}a$ dimer and (c) on a $2a$ dimer. 
The DOS, prior to convolution to obtain the dI/dV is shown in (b) and (d) respectively.
        \label{FigS1}}
\end{figure}

We evaluate the local density of states to approximate the STM data.  The calculations are first performed for a single Cr atom. We find that the experimental spectra (position and amplitude of Shiba excitations) are best fitted for a exchange coupling  $J$  of 1.8 eV  and a scattering potential $K$ of  5.5 eV. In order to compare with the experiment, we further convolute the data by the tip's and sample's BCS DOS. This leads to the appearance of the QP coherent peak that is otherwise absent of the DOS on the Cr atoms.  From a typical spectrum on the bare substrate we fit an  effective Dynes damping of 20 $\mu$eV and a tip gap of $\Delta\approx$ 700~$\mu$eV. 
In the simulations we also added Gaussian broadening of 70~$\mu$eV  to take into account different sources of noise (AC voltage from the lock-in amplifier and superconducting gap distribution). With these parameters we reproduce the dI/dV
spectrum of a single Cr atom. 

Next, we evaluate the full set of data for the dimers without
any extra parameters, only the distance between Cr atoms and the relative orientation of
the atomic magnetic moments characterize the dimers.
Figure~\ref{FigS1}a and \ref{FigS1}b  show the DOS and dI/dV spectrum  for the $\sqrt{2}a$ dimer, respectively. 
The plots show both the simulations imposing parallel (FM) and antiparallel (AFM) Cr magnetic moments. 
We see that the splitting observed in the experiments is retrieved for the FM case.  Figure~\ref{FigS1}c and \ref{FigS1}d shows similar results for the $2a$ dimer. Note that in this model there are no atomistic details. Threfore, the $2a$ dimer simply corresponds to a dimer with larger interatomic distance. In this case, the single pair of Shiba excitations is observed for the AFM alignment, instead. 

%The simulation of the $2a$ dimer also reveals an assymetric behaviour for the FM case: while a negative bias a split pair of  Shiba excitations is observed, the corresponding particle ones collapse into one same peak.  The origin of this peculiar behavior is the presence of a finite potential scattering, which not only imposes an asymmetry in the amplitude of particle and hole peaks, but also affects the split of particle and hole excitations differently.   Figure~\ref{FigS12} shows the behavior of the position of particle and hole peaks with the dimer separation. For  FM (Fig.~\ref{FigS12}v  ordering,   the particle states coalesce   at d=$2a$ and separate again at larger distances. The AFM curves (Fig.~\ref{FigS12}a are more simple, showing a single excitation for hoelas and particles for every distance, in agreement with the  behavior obseved for the $1a$ and $2a$. 

%-----------------%
% Figure (Shiba)
%-----------------%
%\begin{figure}[t]
%        \begin{center}
%                \includegraphics[width=.95\linewidth]{figureS12}
%        \end{center}
%        \vspace{-0.3cm}
%        \caption{
%Behaviour of the Shiba states with interatomic (Cr--Cr) distance for AFM (a) and FM (b) dimers.
%Particle states are plotted in black (above the Fermi energy, which is the zero of energies)
%and the hole states are plotted in red.
%        \label{FigS12}}
%\end{figure}
%-----------------% 

\newpage
\subsection{Note 4 - DFT simulations}

To better understand and interpret our experimental findings, we performed DFT calculations on this system.
The calculations were performed using
the VASP code~\cite{vasp}. The \BiPd\ slab was optimized using the
Perdew-Burke-Ernzerhof (PBE) form of the generalised gradient
approximation (GGA)~\cite{PBE}, obtaining a bulk lattice parameter
$a=b=3.406$~\AA~and $c=13.011$~\AA~in good agreement with other
DFT calculations~\cite{Shein_2013} and the experimental value
of $3.36(8)$~\AA~and $12.97(2)$~\AA~given in Ref.~\cite{Herrera_2015}.
The  surface calculations were performed for Bi-terminated slabs with
four Bi layers and two Pd ones. The surface unit cell was taken as a
$6\times4$ lattice, where two Cr atoms can be placed at $2 a$ without
interaction between dimers. The k-point sampling was $1\times 3 \times 1$.
The structures were relaxed until forces were smaller than 0.01
eV/\AA~for the three topmost layers and the Cr structures. The magnetic
configurations were evaluated by fixing the collinear atomic magnetic
moments, breaking spin symmetry.

First we calculated the stability of a single Cr atoms adsorbed on different positions of the $\beta$-Bi$_2$Pd surface. 
The calculations show that Cr adatoms preferentially adsorb
on the hollow site where one Pd atom is directly underneath.
As a result of the interaction, the Pd atom shifts
by 0.1~\AA~towards the Cr away from the Pd layer. The hollow site is
preferred with an adsorption energy of -2.00 eV per Cr atom, \textit{vs}
-1.45 eV of the bridge site and -1.13 eV of the top site. The total
magnetic moment of the Cr atom goes from 5 $\mu_B$ in the gas phase to
4.4 $\mu_B$ on the hollow site, 4.75 $\mu_B$ on bridge and 4.96 $\mu_B$ on top,
showing the demagnetisation of the atom as it gets more strongly bound to
the surface.  The Cr adatom lies 1.59~\AA~above the Bi surface layer
producing a small expansion of the local Bi-Bi distance to 3.45~\AA~from
the computed Bi-Bi distance of 3.405~\AA.

\nacho{Experimentally, the built structures can be classified by the Cr-Cr
distance in units of the lattice parameter $a$ (see main text).  We calculated the adsorption properties for different types of dimers  by placing atoms in proximal hollow sites, and their respective stability is described in the main text.}

\nacho{We also calculated the absorption of  a compact dimer formed by two Cr atoms on a hollow site, with a Cr-Cr distance of 1.58~\AA, equal to the free-dimer one. In this case, the antiferromagnetic order of the free dimer persist on the surface, resulting in an unpolarized singlet state with negligible magnetic interaction with the surface.   The adsorption energy is 95 meV less favorable than for
the antiferromagnetic $1a$ dimer. So we discard that these structures appear on the  surface. }

\begin{figure}[t]
  	\begin{minipage}[c]{0.60\textwidth}
  	\includegraphics[width=0.99\textwidth]{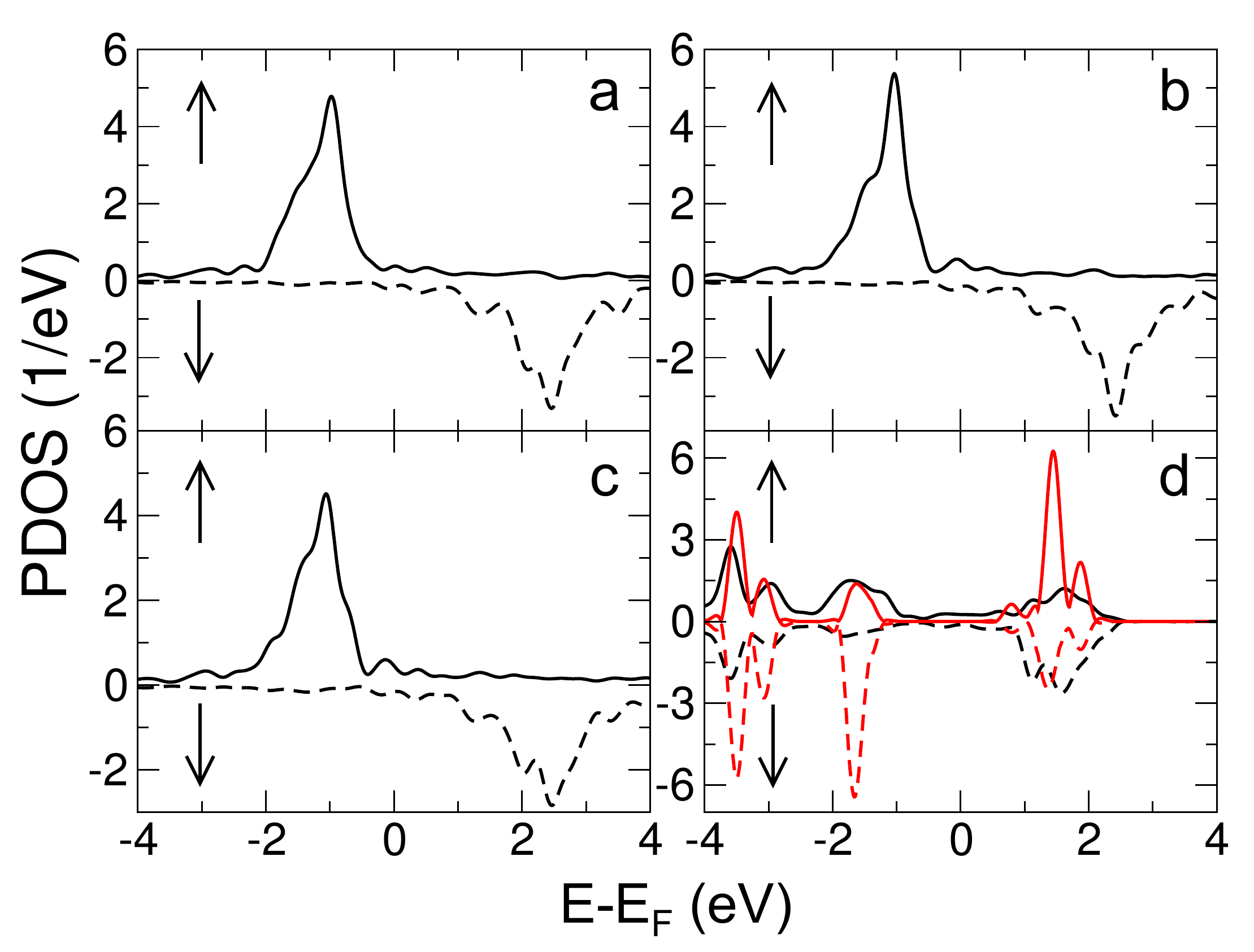}
  \end{minipage}\hfill
         	\begin{minipage}[c]{0.40\textwidth}
        	\caption{\small
        		Density of states projected (PDOS) on a Cr for (a) a single Cr atom
        		adsorbed on the hollow site of the $\beta$-Bi$_2$Pd surface,
        		(b) on one of the Cr atoms of a 2-unit cell apart, $2a$, dimer, the Cr atom is adsorbed on a hollow site,
        		(c) on one of the Cr atoms of a $1a$ dimer, and (d) on
        		one of the Cr atoms of a dimer adsorbed
        		along the surface diagonal centered on the hollow site (black)
        		compared with the free-dimer PDOS (red). Full lines are the PDOS on the majority spin
        		and dashed lines on the minority spin.
        		\label{FigS2}}
        \end{minipage}\hfill
        
\end{figure}

Figure~\ref{FigS2} (a) shows the density of states projected on the
electronic structure of an isolated Cr atom. The strong magnetization of the Cr atom is
preserved upon adsorption as seen by the clear separation of both spins components of the projected DOS. A
slight charge transfer and a sizeable broadening of the atomic levels
reveal a non-negligible interaction with the surface in agreement with
the above values for the atomic adsorption energy and demagnetization.
Figure~\ref{FigS2} (b) shows the density of states projected on the Cr $d$-manifold
of one of the Cr atoms of the $2a$ dimer. The resemblance with
the previous case is notable, showing that the $2a$ dimer in a zero
order approximation can be considered as independent Cr atoms. However
a certain deformation of the PDOS can be seen close to the Fermi energy, implying
a small degree of electronic hybridisation between the two atoms. 
This hybridisation is slightly larger for the $1a$ dimer, Fig. ~\ref{FigS2} (c), 
albeit still very small. Overall, the local environment of the hollow site
controls the main features of the PDOS on the eV scale and thus, the PDOS is largely
equivalent for the three previous cases.

\nacho{As a comparison, we show in Figure~\ref{FigS2} (d) the PDOS on compact Cr$_2$ dimer adsorbed on a hollow site (black lines). The PDOS has no resemblance with the single Cr one. Indeed, the magnetic
moment is much reduced leading to less difference between majority
and minority spin. The same figure shows the PDOS on one of the two Cr atoms of
the gas-phase dimer (red lines), for comparison. We retrieve the same features as for the adsorbed Cr$_2$ compact dimer, except that in the latter case the interaction with
the surface leads to a substantial broadening of the electronic states. }

%-----------------%
% Figure (PDOS)
%-----------------%
%-----------------% 

\nacho{Calculations revealing the spin polarization and ground state magnetic ordering of the systems are shown in the main text. Adsorption of Cr along the troughs of the bismuth top-most layer of $\beta$-Bi$_2$Pd leads to antiferromagnetic spin ordering. Along the diagonal
of the surface, the $\sqrt{2}a$   and the $2\sqrt{2}a$ dimers show ferromagnetic ordering. However, the energy difference to an antiferromagnetic alignment is small (19 and 6 meV respectively),
and further studies are needed to conclude on the actual
DFT values.}

Despite of the large
spin-orbit coupling of the Bi layer, our calculations
with explicit spin-orbit coupling show
that the antiferromagnetic
ordering is favored at short distances along
the surface troughs. However,
calculations with larger unit cells should be performed to be
able to conclude on
more complex non-collinear patterns.

% We have done these for  an infinite chain of Cr atoms separated by d=$1a$, see Fig.~\ref{FigS4} (b-e). In other words, these dimer calculations corresponded to infinite chains with a periodic unit cell with two Cr atoms one unit cell apart.  These calculations clearly favor the antiferromagnetic solution over the other configurations, Fig.~\ref{FigS4} (b). These results seem to indicate that despite of the large spin-orbit coupling of the Bi layer, the antiferromagnetic ordering is favored at short distances. Heavier calculations should be performed to conclude on more complex non-collinear patterns.

%negligible spin-polarisation of the substrate. This result emphasizes the local character of the 
%spin singlet corresponding to the Cr$_2$ dimer.

%Our calculations can be easily extended to infinite chains. We have
%computed infinite chains with Cr atoms every one and two lattice
%parameters.  We find the same type of spin polarisation and the
%prevailance of antiferromagnetic ordering.

\bibliography{dimer_SOM}